\documentclass[a4paper]{article}

\usepackage{romp33}
\usepackage{amsfonts,latexsym%
  ,amsmath,bbm,epsfig%
  }
\makeatletter
 
\def\diagram{\m@th\leftwidth=\z@ \rightwidth=\z@ \topheight=\z@
\botheight=\z@ \setbox\@picbox\hbox\bgroup}
 
\def\enddiagram{\egroup\wd\@picbox\rightwidth\unitlength
\ht\@picbox\topheight\unitlength \dp\@picbox\botheight\unitlength
\hskip\leftwidth\unitlength\box\@picbox}
 
\def\bfig{\begin{diagram}}
\def\efig{\end{diagram}}
\newcount\wideness \newcount\leftwidth \newcount\rightwidth
\newcount\highness \newcount\topheight \newcount\botheight
 
\def\ratchet#1#2{\ifnum#1<#2 \global #1=#2 \fi}
 
\def\putbox(#1,#2)#3{%
\horsize{\wideness}{#3} \divide\wideness by 2
{\advance\wideness by #1 \ratchet{\rightwidth}{\wideness}}
{\advance\wideness by -#1 \ratchet{\leftwidth}{\wideness}}
\vertsize{\highness}{#3} \divide\highness by 2
{\advance\highness by #2 \ratchet{\topheight}{\highness}}
{\advance\highness by -#2 \ratchet{\botheight}{\highness}}
\put(#1,#2){\makebox(0,0){$#3$}}}
 
\def\putlbox(#1,#2)#3{%
\horsize{\wideness}{#3}
{\advance\wideness by #1 \ratchet{\rightwidth}{\wideness}}
{\ratchet{\leftwidth}{-#1}}
\vertsize{\highness}{#3} \divide\highness by 2
{\advance\highness by #2 \ratchet{\topheight}{\highness}}
{\advance\highness by -#2 \ratchet{\botheight}{\highness}}
\put(#1,#2){\makebox(0,0)[l]{$#3$}}}
 
\def\putrbox(#1,#2)#3{%
\horsize{\wideness}{#3}
{\ratchet{\rightwidth}{#1}}
{\advance\wideness by -#1 \ratchet{\leftwidth}{\wideness}}
\vertsize{\highness}{#3} \divide\highness by 2
{\advance\highness by #2 \ratchet{\topheight}{\highness}}
{\advance\highness by -#2 \ratchet{\botheight}{\highness}}
\put(#1,#2){\makebox(0,0)[r]{$#3$}}}

\def\adjust[#1]{} 
 
\newcount \coefa
\newcount \coefb
\newcount \coefc
\newcount\tempcounta
\newcount\tempcountb
\newcount\tempcountc
\newcount\tempcountd
\newcount\xext
\newcount\yext
\newcount\xoff
\newcount\yoff
\newcount\gap%
\newcount\arrowtypea
\newcount\arrowtypeb
\newcount\arrowtypec
\newcount\arrowtyped
\newcount\arrowtypee
\newcount\height
\newcount\width
\newcount\xpos
\newcount\ypos
\newcount\run
\newcount\rise
\newcount\arrowlength
\newcount\halflength
\newcount\arrowtype
\newdimen\tempdimen
\newdimen\xlen
\newdimen\ylen
\newsavebox{\tempboxa}%
\newsavebox{\tempboxb}%
\newsavebox{\tempboxc}%
 
\newdimen\w@dth
 
\def\setw@dth#1#2{\setbox\z@\hbox{\m@th$#1$}\w@dth=\wd\z@
\setbox\@ne\hbox{\m@th$#2$}\ifnum\w@dth<\wd\@ne \w@dth=\wd\@ne \fi
\advance\w@dth by 1.2em}
 
\def\t@^#1_#2{\allowbreak\def\n@one{#1}\def\n@two{#2}\mathrel
{\setw@dth{#1}{#2}
\mathop{\hbox to \w@dth{\rightarrowfill}}\limits
\ifx\n@one\empty\else ^{\box\z@}\fi
\ifx\n@two\empty\else _{\box\@ne}\fi}}
\def\t@@^#1{\@ifnextchar_{\t@^{#1}}{\t@^{#1}_{}}}
\def\to{\@ifnextchar^{\t@@}{\t@@^{}}}
 
\def\t@left^#1_#2{\def\n@one{#1}\def\n@two{#2}\mathrel{\setw@dth{#1}{#2}
\mathop{\hbox to \w@dth{\leftarrowfill}}\limits
\ifx\n@one\empty\else ^{\box\z@}\fi
\ifx\n@two\empty\else _{\box\@ne}\fi}}
\def\t@@left^#1{\@ifnextchar_{\t@left^{#1}}{\t@left^{#1}_{}}}
\def\toleft{\@ifnextchar^{\t@@left}{\t@@left^{}}}
 
\def\two@^#1_#2{\allowbreak
\def\n@one{#1}\def\n@two{#2}\mathrel{\setw@dth{#1}{#2}
\mathop{\vcenter{\lineskip\z@\baselineskip\z@
                 \hbox to \w@dth{\rightarrowfill}%
                 \hbox to \w@dth{\rightarrowfill}}%
       }\limits
\ifx\n@one\empty\else ^{\box\z@}\fi
\ifx\n@two\empty\else _{\box\@ne}\fi}}
\def\tw@@^#1{\@ifnextchar _{\two@^{#1}}{\two@^{#1}_{}}}
\def\two{\@ifnextchar ^{\tw@@}{\tw@@^{}}}
 
\def\tofr@^#1_#2{\def\n@one{#1}\def\n@two{#2}\mathrel{\setw@dth{#1}{#2}
\mathop{\vcenter{\hbox to \w@dth{\rightarrowfill}\kern-1.7ex
                 \hbox to \w@dth{\leftarrowfill}}%
       }\limits
\ifx\n@one\empty\else ^{\box\z@}\fi
\ifx\n@two\empty\else _{\box\@ne}\fi}}
\def\t@fr@^#1{\@ifnextchar_ {\tofr@^{#1}}{\tofr@^{#1}_{}}}
\def\tofro{\@ifnextchar^ {\t@fr@}{\t@fr@^{}}}

\def\mon{\mathop{\m@th\hbox to
      14.6\P@{\lasyb\char'51\hskip-2.1\P@$\arrext$\hss
$\mathord\rightarrow$}}\limits} 
\def\leftmono{\mathrel{\m@th\hbox to
14.6\P@{$\mathord\leftarrow$\hss$\arrext$\hskip-2.1\P@\lasyb\char'50%
}}\limits} 
\mathchardef\arrext="0200       

\def\settypes(#1,#2,#3){\arrowtypea#1 \arrowtypeb#2 \arrowtypec#3}
\def\settoheight#1#2{\setbox\@tempboxa\hbox{#2}#1\ht\@tempboxa\relax}%
\def\settodepth#1#2{\setbox\@tempboxa\hbox{#2}#1\dp\@tempboxa\relax}%
\def\settokens`#1`#2`#3`#4`{%
     \def\tokena{#1}\def\tokenb{#2}\def\tokenc{#3}\def\tokend{#4}}
\def\setsqparms[#1`#2`#3`#4;#5`#6]{%
\arrowtypea #1
\arrowtypeb #2
\arrowtypec #3
\arrowtyped #4
\width #5
\height #6
}
\def\setpos(#1,#2){\xpos=#1 \ypos#2}

\def\settriparms[#1`#2`#3;#4]{\settripairparms[#1`#2`#3`1`1;#4]}%
 
\def\settripairparms[#1`#2`#3`#4`#5;#6]{%
\arrowtypea #1
\arrowtypeb #2
\arrowtypec #3
\arrowtyped #4
\arrowtypee #5
\width #6
\height #6
}
 
\def\resetparms{\settripairparms[1`1`1`1`1;500]\width 500}
 
\resetparms
 
\def\mvector(#1,#2)#3{
\put(0,0){\vector(#1,#2){#3}}%
\put(0,0){\vector(#1,#2){26}}%
}
\def\evector(#1,#2)#3{{
\arrowlength #3
\put(0,0){\vector(#1,#2){\arrowlength}}%
\advance \arrowlength by-30
\put(0,0){\vector(#1,#2){\arrowlength}}%
}}
 
\def\horsize#1#2{%
\settowidth{\tempdimen}{$#2$}%
#1=\tempdimen
\divide #1 by\unitlength
}
 
\def\vertsize#1#2{%
\settoheight{\tempdimen}{$#2$}%
#1=\tempdimen
\settodepth{\tempdimen}{$#2$}%
\advance #1 by\tempdimen
\divide #1 by\unitlength
}
 
\def\putvector(#1,#2)(#3,#4)#5#6{{%
\ifnum3<\arrowtype
\putdashvector(#1,#2)(#3,#4)#5\arrowtype
\else
\ifnum\arrowtype<-3
\putdashvector(#1,#2)(#3,#4)#5\arrowtype
\else
\xpos=#1
\ypos=#2
\run=#3
\rise=#4
\arrowlength=#5
\ifnum \arrowtype<0
    \ifnum \run=0
        \advance \ypos by-\arrowlength
    \else
        \tempcounta \arrowlength
        \multiply \tempcounta by\rise
        \divide \tempcounta by\run
        \ifnum\run>0
            \advance \xpos by\arrowlength
            \advance \ypos by\tempcounta
        \else
            \advance \xpos by-\arrowlength
            \advance \ypos by-\tempcounta
        \fi
    \fi
    \multiply \arrowtype by-1
    \multiply \rise by-1
    \multiply \run by-1
\fi
\ifcase \arrowtype
\or \put(\xpos,\ypos){\vector(\run,\rise){\arrowlength}}%
\or \put(\xpos,\ypos){\mvector(\run,\rise)\arrowlength}%
\or \put(\xpos,\ypos){\evector(\run,\rise){\arrowlength}}%
\fi\fi\fi
}}
 
\def\putsplitvector(#1,#2)#3#4{
\xpos #1
\ypos #2
\arrowtype #4
\halflength #3
\arrowlength #3
\gap 140
\advance \halflength by-\gap
\divide \halflength by2
\ifnum\arrowtype>0
   \ifcase \arrowtype
   \or \put(\xpos,\ypos){\line(0,-1){\halflength}}%
       \advance\ypos by-\halflength
       \advance\ypos by-\gap
       \put(\xpos,\ypos){\vector(0,-1){\halflength}}%
   \or \put(\xpos,\ypos){\line(0,-1)\halflength}%
       \put(\xpos,\ypos){\vector(0,-1)3}%
       \advance\ypos by-\halflength
       \advance\ypos by-\gap
       \put(\xpos,\ypos){\vector(0,-1){\halflength}}%
   \or \put(\xpos,\ypos){\line(0,-1)\halflength}%
       \advance\ypos by-\halflength
       \advance\ypos by-\gap
       \put(\xpos,\ypos){\evector(0,-1){\halflength}}%
   \fi
\else \arrowtype=-\arrowtype
   \ifcase\arrowtype
   \or \advance \ypos by-\arrowlength
       \put(\xpos,\ypos){\line(0,1){\halflength}}%
       \advance\ypos by\halflength
       \advance\ypos by\gap
       \put(\xpos,\ypos){\vector(0,1){\halflength}}%
   \or \advance \ypos by-\arrowlength
       \put(\xpos,\ypos){\line(0,1)\halflength}%
       \put(\xpos,\ypos){\vector(0,1)3}%
       \advance\ypos by\halflength
       \advance\ypos by\gap
       \put(\xpos,\ypos){\vector(0,1){\halflength}}%
   \or \advance \ypos by-\arrowlength
       \put(\xpos,\ypos){\line(0,1)\halflength}%
       \advance\ypos by\halflength
       \advance\ypos by\gap
       \put(\xpos,\ypos){\evector(0,1){\halflength}}%
   \fi
\fi
}
 
\def\putmorphism(#1)(#2,#3)[#4`#5`#6]#7#8#9{{%
\run #2
\rise #3
\ifnum\rise=0
  \puthmorphism(#1)[#4`#5`#6]{#7}{#8}#9%
\else\ifnum\run=0
  \putvmorphism(#1)[#4`#5`#6]{#7}{#8}#9%
\else
\setpos(#1)%
\arrowlength #7
\arrowtype #8
\ifnum\run=0
\else\ifnum\rise=0
\else
\ifnum\run>0
    \coefa=1
\else
   \coefa=-1
\fi
\ifnum\arrowtype>0
   \coefb=0
   \coefc=-1
\else
   \coefb=\coefa
   \coefc=1
   \arrowtype=-\arrowtype
\fi
\width=2
\multiply \width by\run
\divide \width by\rise
\ifnum \width<0  \width=-\width\fi
\advance\width by60
\if l#9 \width=-\width\fi
\putbox(\xpos,\ypos){#4}
{\multiply \coefa by\arrowlength
\advance\xpos by\coefa
\multiply \coefa by\rise
\divide \coefa by\run
\advance \ypos by\coefa
\putbox(\xpos,\ypos){#5} }%
{\multiply \coefa by\arrowlength
\divide \coefa by2
\advance \xpos by\coefa
\advance \xpos by\width
\multiply \coefa by\rise
\divide \coefa by\run
\advance \ypos by\coefa
\if l#9%
   \putrbox(\xpos,\ypos){#6}%
\else\if r#9%
   \putlbox(\xpos,\ypos){#6}%
\fi\fi }%
{\multiply \rise by-\coefc
\multiply \run by-\coefc
\multiply \coefb by\arrowlength
\advance \xpos by\coefb
\multiply \coefb by\rise
\divide \coefb by\run
\advance \ypos by\coefb
\multiply \coefc by70
\advance \ypos by\coefc
\multiply \coefc by\run
\divide \coefc by\rise
\advance \xpos by\coefc
\multiply \coefa by140
\multiply \coefa by\run
\divide \coefa by\rise
\advance \arrowlength by\coefa
\ifcase\arrowtype
\or \put(\xpos,\ypos){\vector(\run,\rise){\arrowlength}}%
\or \put(\xpos,\ypos){\mvector(\run,\rise){\arrowlength}}%
\or \put(\xpos,\ypos){\evector(\run,\rise){\arrowlength}}%
\fi}\fi\fi\fi\fi}}

\newcount\numbdashes \newcount\lengthdash \newcount\increment
 
\def\howmanydashes{
\numbdashes=\arrowlength \lengthdash=40
\divide\numbdashes by \lengthdash
\lengthdash=\arrowlength
\divide\lengthdash by \numbdashes
\increment=\lengthdash
\multiply\lengthdash by 3
\divide\lengthdash by 5
}
 
\def\putdashvector(#1)(#2,#3)#4#5{%
\ifnum#3=0 \putdashhvector(#1){#4}#5
\else
\ifnum#2=0
\putdashvvector(#1){#4}#5\fi\fi}
 
\def\putdashhvector(#1,#2)#3#4{{%
\arrowlength=#3 \howmanydashes
\multiput(#1,#2)(\increment,0){\numbdashes}%
{\vrule height .4pt width \lengthdash\unitlength}
\arrowtype=#4 \xpos=#1
\ifnum\arrowtype<0 \advance\arrowtype by 7 \fi
\ifcase\arrowtype
\or \advance\xpos by 10
    \put(\xpos,#2){\vector(-1,0){\lengthdash}}
    \advance\xpos by 40
    \put(\xpos,#2){\vector(-1,0){\lengthdash}}
\or \advance \xpos by 10
    \put(\xpos,#2){\vector(-1,0){\lengthdash}}
    \advance\xpos by  \arrowlength
    \advance\xpos by  -50
    \put(\xpos,#2){\vector(-1,0){\lengthdash}}
\or \advance\xpos by 10
    \put(\xpos,#2){\vector(-1,0){\lengthdash}}
\or \advance\xpos by \arrowlength
    \advance\xpos by -\lengthdash
    \put(\xpos,#2){\vector(1,0){\lengthdash}}
\or {\advance\xpos by 10
    \put(\xpos,#2){\vector(1,0){\lengthdash}}}
    \advance\xpos by \arrowlength
    \advance\xpos by -\lengthdash
    \put(\xpos,#2){\vector(1,0){\lengthdash}}
\or \advance\xpos by \arrowlength
    \advance\xpos by -\lengthdash
    \put(\xpos,#2){\vector(1,0){\lengthdash}}
    \advance\xpos by -40
    \put(\xpos,#2){\vector(1,0){\lengthdash}}
   \fi
}}
 
\def\putdashvvector(#1,#2)#3#4{{%
\arrowlength=#3 \howmanydashes
\ypos=#2 \advance\ypos by -\arrowlength
\multiput(#1,#2)(0,\increment){\numbdashes}%
    {\vrule width .4pt height \lengthdash\unitlength}
\arrowtype=#4 \ypos=#2
\ifnum\arrowtype<0 \advance\arrowtype by 7 \fi
\ifcase\arrowtype
\or \advance\ypos by \arrowlength \advance\ypos by -40
    \put(#1,\ypos){\vector(0,1){\lengthdash}}
    \advance\ypos by -40
    \put(#1,\ypos){\vector(0,1){\lengthdash}}
\or \advance\ypos by 10
    \put(#1,\ypos){\vector(0,1){\lengthdash}}
    \advance\ypos by \arrowlength \advance\ypos by -40
    \put(#1,\ypos){\vector(0,1){\lengthdash}}
\or \advance\ypos by \arrowlength \advance\ypos by -40
    \put(#1,\ypos){\vector(0,1){\lengthdash}}
\or \advance\ypos by 10
    \put(#1,\ypos){\vector(0,-1){\lengthdash}}
\or \advance\ypos by 10
    \put(#1,\ypos){\vector(0,-1){\lengthdash}}
    \advance\ypos by \arrowlength \advance\ypos by -40
    \put(#1,\ypos){\vector(0,-1){\lengthdash}}
\or \advance\ypos by 10
    \put(#1,\ypos){\vector(0,-1){\lengthdash}}
    \advance\ypos by 40
    \put(#1,\ypos){\vector(0,-1){\lengthdash}}
\fi
}}
 
\def\puthmorphism(#1,#2)[#3`#4`#5]#6#7#8{{%
\xpos #1
\ypos #2
\width #6
\arrowlength #6
\arrowtype=#7
\putbox(\xpos,\ypos){#3\vphantom{#4}}%
{\advance \xpos by\arrowlength
\putbox(\xpos,\ypos){\vphantom{#3}#4}}%
\horsize{\tempcounta}{#3}%
\horsize{\tempcountb}{#4}%
\divide \tempcounta by2
\divide \tempcountb by2
\advance \tempcounta by30
\advance \tempcountb by30
\advance \xpos by\tempcounta
\advance \arrowlength by-\tempcounta
\advance \arrowlength by-\tempcountb
\putvector(\xpos,\ypos)(1,0)\arrowlength\arrowtype
\divide \arrowlength by2
\advance \xpos by\arrowlength
\vertsize{\tempcounta}{#5}%
\divide\tempcounta by2
\advance \tempcounta by20
\if a#8 %
   \advance \ypos by\tempcounta
   \putbox(\xpos,\ypos){#5}%
\else
   \advance \ypos by-\tempcounta
   \putbox(\xpos,\ypos){#5}%
\fi}}
 
\def\putvmorphism(#1,#2)[#3`#4`#5]#6#7#8{{%
\xpos #1
\ypos #2
\arrowlength #6
\arrowtype #7
\settowidth{\xlen}{$#5$}%
\putbox(\xpos,\ypos){#3}%
{\advance \ypos by-\arrowlength
\putbox(\xpos,\ypos){#4}}%
{\advance\arrowlength by-140
\advance \ypos by-70
\ifdim\xlen>0pt
   \if m#8%
      \putsplitvector(\xpos,\ypos)\arrowlength\arrowtype
   \else
   \putvector(\xpos,\ypos)(0,-1)\arrowlength\arrowtype
   \fi
\else
   \putvector(\xpos,\ypos)(0,-1)\arrowlength\arrowtype
\fi}%
\ifdim\xlen>0pt
   \divide \arrowlength by2
   \advance\ypos by-\arrowlength
   \if l#8%
      \advance \xpos by-40
      \putrbox(\xpos,\ypos){#5}%
   \else\if r#8%
      \advance \xpos by40
      \putlbox(\xpos,\ypos){#5}%
   \else
      \putbox(\xpos,\ypos){#5}%
   \fi\fi
\fi
}}
 
\def\putsquarep<#1>(#2)[#3;#4`#5`#6`#7]{{%
\setsqparms[#1]%
\setpos(#2)%
\settokens`#3`%
\puthmorphism(\xpos,\ypos)[\tokenc`\tokend`{#7}]{\width}{\arrowtyped}b%
\advance\ypos by \height
\puthmorphism(\xpos,\ypos)[\tokena`\tokenb`{#4}]{\width}{\arrowtypea}a%
\putvmorphism(\xpos,\ypos)[``{#5}]{\height}{\arrowtypeb}l%
\advance\xpos by \width
\putvmorphism(\xpos,\ypos)[``{#6}]{\height}{\arrowtypec}r%
}}
 
\def\putsquare{\@ifnextchar <{\putsquarep}{\putsquarep%
   <\arrowtypea`\arrowtypeb`\arrowtypec`\arrowtyped;\width`\height>}}
\def\square{\@ifnextchar< {\squarep}{\squarep
   <\arrowtypea`\arrowtypeb`\arrowtypec`\arrowtyped;\width`\height>}}
\def\squarep<#1>[#2`#3`#4`#5;#6`#7`#8`#9]{{
\setsqparms[#1]
\diagram
\putsquarep<\arrowtypea`\arrowtypeb`\arrowtypec`
\arrowtyped;\width`\height>
(0,0)[#2`#3`#4`{#5};#6`#7`#8`{#9}]
\enddiagram
}}                                                 
\def\putptrianglep<#1>(#2,#3)[#4`#5`#6;#7`#8`#9]{{%
\settriparms[#1]%
\xpos=#2 \ypos=#3
\advance\ypos by \height
\puthmorphism(\xpos,\ypos)[#4`#5`{#7}]{\height}{\arrowtypea}a%
\putvmorphism(\xpos,\ypos)[`#6`{#8}]{\height}{\arrowtypeb}l%
\advance\xpos by\height
\putmorphism(\xpos,\ypos)(-1,-1)[``{#9}]{\height}{\arrowtypec}r%
}}
 
\def\putptriangle{\@ifnextchar <{\putptrianglep}{\putptrianglep
   <\arrowtypea`\arrowtypeb`\arrowtypec;\height>}}
\def\ptriangle{\@ifnextchar <{\ptrianglep}{\ptrianglep
   <\arrowtypea`\arrowtypeb`\arrowtypec;\height>}}
\def\ptrianglep<#1>[#2`#3`#4;#5`#6`#7]{{
\settriparms[#1]
\diagram
\putptrianglep<\arrowtypea`\arrowtypeb`
\arrowtypec;\height>
(0,0)[#2`#3`#4;#5`#6`{#7}]
\enddiagram
}}                                            
 
\def\putqtrianglep<#1>(#2,#3)[#4`#5`#6;#7`#8`#9]{{%
\settriparms[#1]%
\xpos=#2 \ypos=#3
\advance\ypos by\height
\puthmorphism(\xpos,\ypos)[#4`#5`{#7}]{\height}{\arrowtypea}a%
\putmorphism(\xpos,\ypos)(1,-1)[``{#8}]{\height}{\arrowtypeb}l%
\advance\xpos by\height
\putvmorphism(\xpos,\ypos)[`#6`{#9}]{\height}{\arrowtypec}r%
}}
 
\def\putqtriangle{\@ifnextchar <{\putqtrianglep}{\putqtrianglep
   <\arrowtypea`\arrowtypeb`\arrowtypec;\height>}}
\def\qtriangle{\@ifnextchar <{\qtrianglep}{\qtrianglep
   <\arrowtypea`\arrowtypeb`\arrowtypec;\height>}}
\def\qtrianglep<#1>[#2`#3`#4;#5`#6`#7]{{
\settriparms[#1]
\width=\height                                
\diagram
\putqtrianglep<\arrowtypea`\arrowtypeb`
\arrowtypec;\height>
(0,0)[#2`#3`#4;#5`#6`{#7}]
\enddiagram
}}
 
\def\putdtrianglep<#1>(#2,#3)[#4`#5`#6;#7`#8`#9]{{%
\settriparms[#1]%
\xpos=#2 \ypos=#3
\puthmorphism(\xpos,\ypos)[#5`#6`{#9}]{\height}{\arrowtypec}b%
\advance\xpos by \height \advance\ypos by\height
\putmorphism(\xpos,\ypos)(-1,-1)[``{#7}]{\height}{\arrowtypea}l%
\putvmorphism(\xpos,\ypos)[#4``{#8}]{\height}{\arrowtypeb}r%
}}
 
\def\putdtriangle{\@ifnextchar <{\putdtrianglep}{\putdtrianglep
   <\arrowtypea`\arrowtypeb`\arrowtypec;\height>}}
\def\dtriangle{\@ifnextchar <{\dtrianglep}{\dtrianglep
   <\arrowtypea`\arrowtypeb`\arrowtypec;\height>}}
\def\dtrianglep<#1>[#2`#3`#4;#5`#6`#7]{{
\settriparms[#1]
\width=\height                                
\diagram
\putdtrianglep<\arrowtypea`\arrowtypeb`
\arrowtypec;\height>
(0,0)[#2`#3`#4;#5`#6`{#7}]
\enddiagram
}}
 
\def\putbtrianglep<#1>(#2,#3)[#4`#5`#6;#7`#8`#9]{{%
\settriparms[#1]%
\xpos=#2 \ypos=#3
\puthmorphism(\xpos,\ypos)[#5`#6`{#9}]{\height}{\arrowtypec}b%
\advance\ypos by\height
\putmorphism(\xpos,\ypos)(1,-1)[``{#8}]{\height}{\arrowtypeb}r%
\putvmorphism(\xpos,\ypos)[#4``{#7}]{\height}{\arrowtypea}l%
}}
 
\def\putbtriangle{\@ifnextchar <{\putbtrianglep}{\putbtrianglep
   <\arrowtypea`\arrowtypeb`\arrowtypec;\height>}}
\def\btriangle{\@ifnextchar <{\btrianglep}{\btrianglep
   <\arrowtypea`\arrowtypeb`\arrowtypec;\height>}}
\def\btrianglep<#1>[#2`#3`#4;#5`#6`#7]{{
\settriparms[#1]
\width=\height                               
\diagram
\putbtrianglep<\arrowtypea`\arrowtypeb`
\arrowtypec;\height>
(0,0)[#2`#3`#4;#5`#6`{#7}]
\enddiagram
}}
 
\def\putAtrianglep<#1>(#2,#3)[#4`#5`#6;#7`#8`#9]{{%
\settriparms[#1]%
\xpos=#2 \ypos=#3
{\multiply \height by2
\puthmorphism(\xpos,\ypos)[#5`#6`{#9}]{\height}{\arrowtypec}b}%
\advance\xpos by\height \advance\ypos by\height
\putmorphism(\xpos,\ypos)(-1,-1)[#4``{#7}]{\height}{\arrowtypea}l%
\putmorphism(\xpos,\ypos)(1,-1)[``{#8}]{\height}{\arrowtypeb}r%
}}
 
\def\putAtriangle{\@ifnextchar <{\putAtrianglep}{\putAtrianglep
   <\arrowtypea`\arrowtypeb`\arrowtypec;\height>}}
\def\Atriangle{\@ifnextchar <{\Atrianglep}{\Atrianglep
   <\arrowtypea`\arrowtypeb`\arrowtypec;\height>}}
\def\Atrianglep<#1>[#2`#3`#4;#5`#6`#7]{{
\settriparms[#1]
\width=\height                                     
\diagram
\putAtrianglep<\arrowtypea`\arrowtypeb`
\arrowtypec;\height>
(0,0)[#2`#3`#4;#5`#6`{#7}]
\enddiagram
}}
 
\def\putAtrianglepairp<#1>(#2)[#3;#4`#5`#6`#7`#8]{{%
\settripairparms[#1]%
\setpos(#2)%
\settokens`#3`%
\puthmorphism(\xpos,\ypos)[\tokenb`\tokenc`{#7}]{\height}{\arrowtyped}b%
\advance\xpos by\height
\puthmorphism(\xpos,\ypos)[\phantom{\tokenc}`\tokend`{#8}]%
{\height}{\arrowtypee}b%
\advance\ypos by\height
\putmorphism(\xpos,\ypos)(-1,-1)[\tokena``{#4}]{\height}{\arrowtypea}l%
\putvmorphism(\xpos,\ypos)[``{#5}]{\height}{\arrowtypeb}m%
\putmorphism(\xpos,\ypos)(1,-1)[``{#6}]{\height}{\arrowtypec}r%
}}
 
\def\putAtrianglepair{\@ifnextchar <{\putAtrianglepairp}{\putAtrianglepairp%
   <\arrowtypea`\arrowtypeb`\arrowtypec`\arrowtyped`\arrowtypee;\height>}}
\def\Atrianglepair{\@ifnextchar <{\Atrianglepairp}{\Atrianglepairp%
   <\arrowtypea`\arrowtypeb`\arrowtypec`\arrowtyped`\arrowtypee;\height>}}
 
\def\Atrianglepairp<#1>[#2;#3`#4`#5`#6`#7]{{
\settripairparms[#1]
\settokens`#2`
\width=\height                                
\diagram
\putAtrianglepairp                            
<\arrowtypea`\arrowtypeb`\arrowtypec`
\arrowtyped`\arrowtypee;\height>
(0,0)[{#2};#3`#4`#5`#6`{#7}]
\enddiagram
}}
 
\def\putVtrianglep<#1>(#2,#3)[#4`#5`#6;#7`#8`#9]{{%
\settriparms[#1]%
\xpos=#2 \ypos=#3
\advance\ypos by\height
{\multiply\height by2
\puthmorphism(\xpos,\ypos)[#4`#5`{#7}]{\height}{\arrowtypea}a}%
\putmorphism(\xpos,\ypos)(1,-1)[`#6`{#8}]{\height}{\arrowtypeb}l%
\advance\xpos by\height
\advance\xpos by\height
\putmorphism(\xpos,\ypos)(-1,-1)[``{#9}]{\height}{\arrowtypec}r%
}}
 
\def\putVtriangle{\@ifnextchar <{\putVtrianglep}{\putVtrianglep
   <\arrowtypea`\arrowtypeb`\arrowtypec;\height>}}
\def\Vtriangle{\@ifnextchar <{\Vtrianglep}{\Vtrianglep
   <\arrowtypea`\arrowtypeb`\arrowtypec;\height>}}
\def\Vtrianglep<#1>[#2`#3`#4;#5`#6`#7]{{
\settriparms[#1]
\width=\height                                 
\diagram
\putVtrianglep<\arrowtypea`\arrowtypeb`
\arrowtypec;\height>
(0,0)[#2`#3`#4;#5`#6`{#7}]
\enddiagram
}}
 
\def\putVtrianglepairp<#1>(#2)[#3;#4`#5`#6`#7`#8]{{
\settripairparms[#1]%
\setpos(#2)%
\settokens`#3`%
\advance\ypos by\height
\putmorphism(\xpos,\ypos)(1,-1)[`\tokend`{#6}]{\height}{\arrowtypec}l%
\puthmorphism(\xpos,\ypos)[\tokena`\tokenb`{#4}]{\height}{\arrowtypea}a%
\advance\xpos by\height
\puthmorphism(\xpos,\ypos)[\phantom{\tokenb}`\tokenc`{#5}]%
{\height}{\arrowtypeb}a%
\putvmorphism(\xpos,\ypos)[``{#7}]{\height}{\arrowtyped}m%
\advance\xpos by\height
\putmorphism(\xpos,\ypos)(-1,-1)[``{#8}]{\height}{\arrowtypee}r%
}}
 
\def\putVtrianglepair{\@ifnextchar <{\putVtrianglepairp}{\putVtrianglepairp%
    <\arrowtypea`\arrowtypeb`\arrowtypec`\arrowtyped`\arrowtypee;\height>}}
\def\Vtrianglepair{\@ifnextchar <{\Vtrianglepairp}{\Vtrianglepairp%
    <\arrowtypea`\arrowtypeb`\arrowtypec`\arrowtyped`\arrowtypee;\height>}}
\def\Vtrianglepairp<#1>[#2;#3`#4`#5`#6`#7]{{
\settripairparms[#1]
\settokens`#2`
\diagram
\putVtrianglepairp                             
<\arrowtypea`\arrowtypeb`\arrowtypec`
\arrowtyped`\arrowtypee;\height>
(0,0)[{#2};#3`#4`#5`#6`{#7}]
\enddiagram
}}

\def\putCtrianglep<#1>(#2,#3)[#4`#5`#6;#7`#8`#9]{{%
\settriparms[#1]%
\xpos=#2 \ypos=#3
\advance\ypos by\height
\putmorphism(\xpos,\ypos)(1,-1)[``{#9}]{\height}{\arrowtypec}l%
\advance\xpos by\height
\advance\ypos by\height
\putmorphism(\xpos,\ypos)(-1,-1)[#4`#5`{#7}]{\height}{\arrowtypea}l%
{\multiply\height by 2
\putvmorphism(\xpos,\ypos)[`#6`{#8}]{\height}{\arrowtypeb}r}%
}}
 
\def\putCtriangle{\@ifnextchar <{\putCtrianglep}{\putCtrianglep
    <\arrowtypea`\arrowtypeb`\arrowtypec;\height>}}
\def\Ctriangle{\@ifnextchar <{\Ctrianglep}{\Ctrianglep
    <\arrowtypea`\arrowtypeb`\arrowtypec;\height>}}
\def\Ctrianglep<#1>[#2`#3`#4;#5`#6`#7]{{
\settriparms[#1]
\width=\height                               
\diagram
\putCtrianglep<\arrowtypea`\arrowtypeb`
\arrowtypec;\height>
(0,0)[#2`#3`#4;#5`#6`{#7}]
\enddiagram
}}                                           
\def\putDtrianglep<#1>(#2,#3)[#4`#5`#6;#7`#8`#9]{{%
\settriparms[#1]%
\xpos=#2 \ypos=#3
\advance\xpos by\height \advance\ypos by\height
\putmorphism(\xpos,\ypos)(-1,-1)[``{#9}]{\height}{\arrowtypec}r%
\advance\xpos by-\height \advance\ypos by\height
\putmorphism(\xpos,\ypos)(1,-1)[`#5`{#8}]{\height}{\arrowtypeb}r%
{\multiply\height by 2
\putvmorphism(\xpos,\ypos)[#4`#6`{#7}]{\height}{\arrowtypea}l}%
}}
 
\def\putDtriangle{\@ifnextchar <{\putDtrianglep}{\putDtrianglep
    <\arrowtypea`\arrowtypeb`\arrowtypec;\height>}}
\def\Dtriangle{\@ifnextchar <{\Dtrianglep}{\Dtrianglep
   <\arrowtypea`\arrowtypeb`\arrowtypec;\height>}}
\def\Dtrianglep<#1>[#2`#3`#4;#5`#6`#7]{{
\settriparms[#1]
\width=\height                              
\diagram
\putDtrianglep<\arrowtypea`\arrowtypeb`
\arrowtypec;\height>
(0,0)[#2`#3`#4;#5`#6`{#7}]
\enddiagram
}}                                          
\def\setrecparms[#1`#2]{\width=#1 \height=#2}%
 
\def\recursep<#1`#2>[#3;#4`#5`#6`#7`#8]{{\m@th
\width=#1 \height=#2
\settokens`#3`
\settowidth{\tempdimen}{$\tokena$}
\ifdim\tempdimen=0pt
  \savebox{\tempboxa}{\hbox{$\tokenb$}}%
  \savebox{\tempboxb}{\hbox{$\tokend$}}%
  \savebox{\tempboxc}{\hbox{$#6$}}%
\else
  \savebox{\tempboxa}{\hbox{$\hbox{$\tokena$}\times\hbox{$\tokenb$}$}}%
  \savebox{\tempboxb}{\hbox{$\hbox{$\tokena$}\times\hbox{$\tokend$}$}}%
  \savebox{\tempboxc}{\hbox{$\hbox{$\tokena$}\times\hbox{$#6$}$}}%
\fi
\ypos=\height
\divide\ypos by 2
\xpos=\ypos
\advance\xpos by \width
\bfig
\putCtrianglep<-1`1`1;\ypos>(0,0)[`\tokenc`;#5`#6`{#7}]%
\puthmorphism(\ypos,0)[\tokend`\usebox{\tempboxb}`{#8}]{\width}{-1}b%
\puthmorphism(\ypos,\height)[\tokenb`\usebox{\tempboxa}`{#4}]{\width}{-1}a%
\advance\ypos by \width
\putvmorphism(\ypos,\height)[``\usebox{\tempboxc}]{\height}1r%
\efig
}}
 
\def\recurse{\@ifnextchar <{\recursep}{\recursep<\width`\height>}}
 
\def\puttwohmorphisms(#1,#2)[#3`#4;#5`#6]#7#8#9{{%
\puthmorphism(#1,#2)[#3`#4`]{#7}0a
\ypos=#2
\advance\ypos by 20
\puthmorphism(#1,\ypos)[\phantom{#3}`\phantom{#4}`#5]{#7}{#8}a
\advance\ypos by -40
\puthmorphism(#1,\ypos)[\phantom{#3}`\phantom{#4}`#6]{#7}{#9}b
}}
 
\def\puttwovmorphisms(#1,#2)[#3`#4;#5`#6]#7#8#9{{%
\putvmorphism(#1,#2)[#3`#4`]{#7}0a
\xpos=#1
\advance\xpos by -20
\putvmorphism(\xpos,#2)[\phantom{#3}`\phantom{#4}`#5]{#7}{#8}l
\advance\xpos by 40
\putvmorphism(\xpos,#2)[\phantom{#3}`\phantom{#4}`#6]{#7}{#9}r
}}
 
\def\puthcoequalizer(#1)[#2`#3`#4;#5`#6`#7]#8#9{{%
\setpos(#1)%
\puttwohmorphisms(\xpos,\ypos)[#2`#3;#5`#6]{#8}11%
\advance\xpos by #8
\puthmorphism(\xpos,\ypos)[\phantom{#3}`#4`#7]{#8}1{#9}
}}
 
\def\putvcoequalizer(#1)[#2`#3`#4;#5`#6`#7]#8#9{{%
\setpos(#1)%
\puttwovmorphisms(\xpos,\ypos)[#2`#3;#5`#6]{#8}11%
\advance\ypos by -#8
\putvmorphism(\xpos,\ypos)[\phantom{#3}`#4`#7]{#8}1{#9}
}}
 
\def\putthreehmorphisms(#1)[#2`#3;#4`#5`#6]#7(#8)#9{{%
\setpos(#1) \settypes(#8)
\if a#9 %
     \vertsize{\tempcounta}{#5}%
     \vertsize{\tempcountb}{#6}%
     \ifnum \tempcounta<\tempcountb \tempcounta=\tempcountb \fi
\else
     \vertsize{\tempcounta}{#4}%
     \vertsize{\tempcountb}{#5}%
     \ifnum \tempcounta<\tempcountb \tempcounta=\tempcountb \fi
\fi
\advance \tempcounta by 60
\puthmorphism(\xpos,\ypos)[#2`#3`#5]{#7}{\arrowtypeb}{#9}
\advance\ypos by \tempcounta
\puthmorphism(\xpos,\ypos)[\phantom{#2}`\phantom{#3}`#4]{#7}{\arrowtypea}{#9}
\advance\ypos by -\tempcounta \advance\ypos by -\tempcounta
\puthmorphism(\xpos,\ypos)[\phantom{#2}`\phantom{#3}`#6]{#7}{\arrowtypec}{#9}
}}
 
\def\setarrowtoks[#1`#2`#3`#4`#5`#6]{%
\def\toka{#1}
\def\tokb{#2}
\def\tokc{#3}
\def\tokd{#4}
\def\toke{#5}
\def\tokf{#6}
}
\def\hex{\@ifnextchar <{\hexp}{\hexp<1000`400>}}
\def\hexp<#1`#2>[#3`#4`#5`#6`#7`#8;#9]{%
\setarrowtoks[#9]
\yext=#2 \advance \yext by #2
\xext=#1 \advance\xext by \yext
\bfig
\putCtriangle<-1`0`1;#2>(0,0)[`#5`;\tokb``\tokd]
\xext=#1 \yext=#2 \advance \yext by #2
\putsquare<1`0`0`1;\xext`\yext>(#2,0)[#3`#4`#7`#8;\toka```\tokf]
\advance \xext by #2
\putDtriangle<0`1`-1;#2>(\xext,0)[`#6`;`\tokc`\toke]
\efig
}
\makeatother

\begin{document}

\title{
De Donder-Weyl Equations and Multisymplectic Geometry%
}

\author{Cornelius Pauf\/ler\thanks{\hspace*{1ex}
    e-mail: paufler@physik.uni-freiburg.de.}
~and
Hartmann R{\"o}mer\thanks{\hspace*{1ex}
  Talk given by H. R. at the 33rd Symposium on
  Mathematical Physics, Toru{\'n}, Poland, June 2001.
  e-mail: roemer@physik.uni-freiburg.de.}\\
Fakult\"at f\"ur Physik\\
Albert-Ludwigs-Universit\"at Freiburg im Breisgau\\
Hermann-Herder-Stra\ss e 3\\
D 79104 Freiburg i. Br.
}
\date{August 31, 2001}

\maketitle

\begin{abstract}
Multisymplectic geometry is an adequate formalism to geometrically describe
first order classical field theories.
The De Donder-Weyl equations are treated in the framework of
multisymplectic geometry, solutions are identified as integral
manifolds of Hamiltonean multivectorfields.\\
In contrast to mechanics, solutions cannot be described by points in
the multisymplectic phase space. Foliations of the configuration space
by solutions and a multisymplectic version of Hamilton-Jacobi theory
are also discussed.
\end{abstract}


The field of multisymplectic geometry has experienced a revival of
active research recently with the discovery of super
Poisson-Lie brackets on the canonically arising multisymplectic phase
spaces (\cite{Kanatchikov:1997,ForgerRoemer:2000}).
It provides a geometrical framework to formulate
classical field theory in a coordinate free manner on arbitrary
space-time manifolds. For a detailed review we refer to the excellent
exposition (\cite{GotayIsenbergMarsden:1998}), here we only sketch the
properties needed for our work.

Consider a classical field $\varphi^i(x)$, $(i=1,\ldots,N)$ over
$n$-dimensional space-time with a Lagrangean density
$\mathcal L(x^\mu,\varphi^i,\partial_\mu\varphi^i)$. The corresponding equations
of motion are
\begin{equation}
  \label{Euler-Lagrange}
  \frac{\partial\mathcal L}{\partial\varphi^i}-
  \partial_\mu\frac{\partial\mathcal L}{\partial(\partial_\mu\varphi^i)}=0.
\end{equation}
The main idea of the multisymplectic formulation of classical field
theory consists in associating $n$ polymomenta
\begin{equation}
  \label{Polymomenta}
  \pi^\mu_i=\frac{\partial\mathcal L}
  {\partial(\partial_\mu\varphi^i)},
  \quad (\mu=1,\ldots,n;\; i=1,\ldots,N)
\end{equation}
to every component of the field  as opposed to the conventional
canonical approach, where only $\pi^0_i=\pi_i$ is considered.\\
The multisymplectic phase space $\mathcal P$, thus arising, has
dimension $(N+1)(n+1)$ and can be described by local coordinates
\begin{equation}
  \label{Coordinates}
  (x^\mu,q^i,p^\mu_i,p),
\end{equation}
where $x^\mu$ correspond to space-time, $q^i$ and $p^\mu_i$ to the
values of the field and its polymomenta, respectively. The importance
of the additional coordinate $p$, necessary for consistency, will be
explained in a moment.\\
If we assume that the Lagrangean $\mathcal L$ be non singular, such that
(\ref{Polymomenta}) can be inverted and the field derivatives
$\partial_\mu\varphi^i$ can be expressed as functions of $x^\mu$,
$\varphi^i$, and $\pi_i^\mu$, then we can define a De Donder-Weyl
Hamiltonean
\begin{equation}
  \label{DWHamiltonean}
  \mathcal
  H(x^\mu,\varphi^i,\pi^\mu_i)
  =\pi^\mu_i\partial_\mu\varphi^i-\mathcal L
\end{equation}
and we readily see that the equations of motion (\ref{Euler-Lagrange})
are equivalent to the De Donder-Weyl equations
\begin{equation}
  \label{DWeq}
  \frac{\partial\mathcal H}{\partial\varphi^i}=-\partial_\mu\pi^\mu_i;
  \quad
  \frac{\partial \mathcal H}{\partial\pi^\mu_i}=\partial_\mu\varphi^i.
\end{equation}
These equations also follow from the variational
principle
\begin{equation}
  \label{Variation}
  \delta\int\,d^nx\,(\pi^\mu_i\,\partial_\mu\varphi^i-\mathcal H)
  =\delta\int\Theta\stackrel!=0,
\end{equation}
where the variation has to be performed in $\varphi$ and $\pi$
independently.
\\
In a geometric formulation, the fields are sections of a bundle
$\mathcal E\stackrel\pi\rightarrow\mathcal M$ over an $n$-dimensional
space-time manifold $\mathcal M$ with typical fibre $\mathcal Q$.\\
The first jet bundle (for an introduction we refer to Saunder's book
\cite{Saunders:1989}) of sections of $\mathcal E$, denoted $\mathfrak
J_1(\mathcal E)$, is an affine bundle over $\mathcal E$, its dual $\mathcal
P=\mathfrak J_1^\ast(\mathcal E)$ consisting of all affine maps into the
line bundle $\Lambda^n TM$ is a vector bundle over
$\mathcal E$. The fibre dimension of $\mathcal P$ equals the fibre dimension of
$\mathfrak J_1(\mathcal E)$ plus $1$, thus accommodating the translation
part of the affine maps. The coordinate $p$, mentioned above, is
associated to these translations.\\
$\mathcal P$ is just the multisymplectic phase space. The
correspondences between classical mechanics and multisymplectic
classical field theory are collected in the following table.
\begin{center}
\begin{tabular}{p{17em}|p{17em}}
  Mechanics ($n=1$)&Field theory\\\hline
  Extended configuration space
  $$\mathcal Q\times \mathbbm R$$
  &
  Bundle over an $n$-dimensional space-time manifold $\mathcal M$ with typical fibre $Q$
  $$\mathcal E\stackrel\pi\rightarrow\mathcal M$$
  \\
  \hline
  Double-extended phase space
  $$\mathcal P=T^\ast(\mathcal Q\times \mathbbm R)$$
  &
  Multisymplectic phase space, affine dual of jet bundle
  $$\mathcal P=\mathfrak J_1^\ast(\mathcal E)$$
  \\
  \hline
  Coordinates of $\mathcal Q\times\mathbbm R$
  $$t,q^i$$
  &
  Coordinates of $\mathcal E$
  $$x_\mu,q^i$$
  \\\hline
  Coordinates of $T^\ast\mathcal Q\times\mathbbm R^2=\mathcal P$
  $$t,q^i,E,p_i$$
  &
  Coordinates of $\mathcal P=\mathfrak J^\ast_1(\mathcal E)$
  $$x^\mu,q^i,p,p^\mu_i$$
  \\\hline
  Poincar\'e-Cartan $1$-form on $\mathcal P$
  $$\Theta=p_i\,dq^i-E\,dt$$
  &
  Poincar\'e-Cartan $n$-form on $\mathcal P$,
  $2$-horizontal\footnote{In this article, horizontal means that the
    form under consideration vanishes upon contraction with a vector
    {\em which is vertical w.r.t. the repsective projection onto the base
    manifold $\mathcal
    M$}. $r$-horizontal then means that it vanishes on every $2$
    vertical vectors.}
  $$\Theta=p_i^\mu\,dq^i \wedge d^nx_\mu-p\,dt\wedge d^nx$$
  \\\hline
  Symplectic $2$-form on $\mathcal P$, non-degenerate
  $$\omega=-d\Theta=dq^i\wedge dp_i+dE\wedge dt$$
  &Multisymplectic $(n+1)$-form on $\mathcal P$, non-degenerate
  (on vectorfields)
  $$\omega=-d\Theta=dq^i\wedge dp_i^\mu\wedge d^nx_\mu
  +dp\wedge d^nx$$
  \\\hline
\end{tabular}
\end{center}
Here, $d^nx=dx^1\wedge\cdots\wedge dx^n$ is the volume form on $\mathcal M$,
and $d^nx_\mu=i_{\partial_\mu}d^nx$.\\
Some remarks are in order here.
\begin{enumerate}
\item Although the forms $\Theta$ and $\omega$ are given in
  coordinates they can be defined intrinsically.
\item It can be read off from its coordinate expression that $\Theta$
  is the most general $2$-horizontal $n$-form on $\mathcal E$
  and that $\mathcal P$ can
  also be identified with the space of all forms of this type.
\item Truncating the translation part by a condition like
  $p=-\mathcal H(x^\mu,q^i,p^\mu_i)$ leeds to a smaller phase space
  $\tilde{\mathcal P}$ of dimension $(N+1)(n+1)-1$.
\end{enumerate}
The relationships between the various spaces and bundle structures in
multisymplectic geometry can be read off the diagram
\begin{equation}
  \label{Diagramm}
  \begin{picture}(2100,1200)
    \putmorphism(1000,500)(0,-1)[\mathcal E`\mathcal M`{}]{500}{1}{l}
    \putmorphism(2000,1000)(-1,-1)[\frak J_1(\mathcal E)`
         \phantom{\mathcal M}`
         {}]{1000}{1}{r}
    \putmorphism(2000,1000)(-2,-1)[\phantom{\frak J^1(\mathcal E)}`
         \phantom{\mathcal M}`
    ]{1000}{1}{r}
    \putmorphism(0,1000)(1,-1)[\tilde{\mathcal P}`
         \phantom{\mathcal M}`
    ]{1000}{1}{r}
    \putmorphism(1000,1000)(1,0)[\mathcal P=\mathfrak J_1^\ast(\mathcal E)`
         \phantom{\frak J_1(\mathcal E)}`
         {\scriptstyle \mathbbm F\mathcal L}
    ]{1000}{-1}{u}
    \putmorphism(0,1000)(1,0)[\phantom{\tilde{\mathcal P}}`\phantom{\mathcal P=\mathfrak J_1^\ast(\mathcal E)}`
    ]{1000}{-1}{r}
    \putmorphism(0,1000)(2,-1)[\phantom{\tilde{\mathcal P}}`\phantom{\mathcal E}`
    ]{1000}{1}{l}
    \putmorphism(1020,1000)(0,1)[\phantom{\mathcal P=\mathfrak J_1^\ast(\mathcal E)}`
         \phantom{\mathcal E}`\pi_{\mathcal E\mathcal P}
    ]{500}{1}{r}
    \putmorphism(980,1000)(0,1)[\phantom{\mathcal P=\mathfrak J_1^\ast(\mathcal E)}`
         \phantom{\mathcal E}`T
    ]{500}{-1}{l}
  \end{picture}
\end{equation}
Here $\mathbbm F\mathcal L$ is the fibre derivative associated to the
Lagrangean density $\mathcal L$. The latter should be interpreted as a
function on the first jet bundle to $\mathcal E$,
\begin{equation}
  \label{LagrangeDens}
  \begin{split}
    \mathcal L:\quad\mathfrak J_1(\mathcal E)&\rightarrow \mathbbm R\\
    (x^\mu,q^i,q^i_\mu)&\mapsto\mathcal L(x^\mu,q^i,q^i_\mu).
  \end{split}
\end{equation}
Then $\mathbbm F\mathcal L$ -- which is also known as the covariant
Legendre transformation -- is given by
\begin{equation}
  \label{Legendre}
  \mathbbm F\mathcal L(x^\mu,q^i,q^i_\mu)
  =(x^\mu,q^i,p^\mu_i=\frac{\partial\mathcal L}{\partial q^i_\mu},
  p=\mathcal L-\frac{\partial L}{\partial q^i\mu}\,q^i_\mu).
\end{equation}
The subject of this note is a description of the De Donder-Weyl
equations and their properties in terms of general multisymplectic
geometry, more precisely in terms of distributions
associated to separable Hamiltonean $n$-vectorfields on $\mathcal P$
and their integral manifolds.\\
In contrast to this programme Echeverr\'{\i}a-Enr\'{\i}qez et
al. (\cite{Echeverria-EnriquezMunoz-LecandaRoman-Roy:1999}) have
worked on the truncated space $\tilde{\mathcal P}$, also employed by
Kanatchikov (\cite{Kanatchikov:1997,Kanatchikov:1998}).
The price to be payed for this is the necessity to
introduce a connection into a fully geometric treatment.\\
Martin (\cite{Martin:1988b}) has investigated
De Donder-Weyl equations on very special
multisymplectic manifolds only. Although those cases show very
interesting additional properties as, for example, a generalised
Darboux theorem (\cite{Martin:1988a}),
they only cover the case when $\mathcal E$ is some
antisymmetrised tensor bundle of the tangent bundle to $\mathcal M$.\\
Let us briefly define the necessary geometric notions. An
{\bf $r$-vectorfield} $X$ on $\mathcal P$ is just a totally
antisymmetric covariant vectorfield on $\mathcal P$, i.e. $X$ is a
section of the $r$-th exterior power of the tangent bundle $T\mathcal P$,
\begin{equation}
  \label{vectorfield}
  X\in\Gamma\Lambda^r T\mathcal P.
\end{equation}
An $r$-vectorfield $X$ is called (locally) {\bf separable}, if there are
vectorfields $Z_1,\ldots,Z_r$ such that (locally)
\begin{equation}
  \label{separable}
  X= Z_1\wedge \cdots \wedge Z_r.
\end{equation}
An $r$-distribution on $\mathcal P$ is a smooth collection of
$r$-dimensional subspaces of the tangent space $T_p\mathcal P$ for all
$p\in\mathcal P$. Taking vectorfields $Z_1,\ldots,Z_r$ whose values
(locally) span the $r$-dimensional subspaces at every point, wee see
that distributions can be described by separable $r$-vectorfields
(\cite{Echeverria-EnriquezMunoz-LecandaRoman-Roy:1999,PauflerRoemer:2001},
for an introduction into the theory of integrable distributions and
foliations see \cite{KolarMichorSlovak:1993}).\\
An $r$-distribution on $\mathcal P$ is called {\bf integrable},
if locally every point $p$ of $\mathcal P$ lies in a unique integral
submanifold $\mathcal N$ of dimension $r$ of $\mathcal P$ such that
for $p\in\mathcal M$ $T_p\mathcal N$ is the subspace of
$T_p\mathcal P$ defined by the distribution. A separable
$r$-vectorfield defining an integrable distribution will also be
called integrable.\\
An $(n-r)$-form $f$ and an $r$-vectorfield $X_f$ are called
{\bf Hamiltonean}, if
\begin{equation}
  \label{HamVf-HamForm}
  df=i_{X_f}\omega.
\end{equation}
There are two essential features that do not occur in the case $n=1$
of classical mechanics.\\
Firstly, although $\omega$ is non-degenerate, $X_f$ is not uniquely defined by
$df$ for $r\neq 1$. \\
Moreover, by far not every $(n-r)$-form $f$ is Hamiltonean. There are in
general strong constraints both on $f$ and on $X_f$ which are
satisfied trivially for $n=1$ (in this case, $r=1$ is the only
possibility) or $r=n$.

Consider now a function $h(x^\mu,q^i,p^\mu_i,p)$ which automatically
gives a Hamiltonean zero form. It turns out that separable Hamiltonean
$n$-vectorfields $X_h$ on $\mathcal P$ with
\begin{equation}\label{defnrel}
  i_{X_h}\omega= dh
\end{equation}
can be found provided $h$ depends on $p$ in a particular way, namely
\begin{equation}
  \label{HamFkt}
  h(x^\mu,q^i,p^\mu_i,p)=-\mathcal H(x^\mu,q^i,p^\mu_i)-p,
\end{equation}
where $\mathcal H$ is a function not depending on $p$.\\
We shall show that the Hamiltonean $n$-vectorfield $X_h$ describes a
solution of the De Donder-Weyl equations (\ref{DWeq}) for $\mathcal H$.
For the solutions of the De Donder-Weyl equations we shall always have
to assume that the projection of the values of $X_h$
down to $\mathcal M$ has to be
of maximal rank. This motivates the ansatz
\begin{equation}
  \label{Ansatz}
  Z_\mu={\textstyle\frac\partial{\partial x^\mu}}
  +(Z_\mu)^i{\textstyle\frac\partial{\partial q^i}}
  +(Z_\mu)^\nu_i{\textstyle\frac\partial{\partial p^\nu_i}}
  +(Z_\mu)_0{\textstyle\frac\partial{\partial p}},\quad
  \mu=1,\ldots,n
\end{equation}
for the $n$ vectorfields that shall be combined to yield $X_h$,
\begin{equation}
  \label{Composition}
  Z_1\wedge\cdots\wedge Z_n\stackrel!=X_h.
\end{equation}
Inserting (\ref{Ansatz}) and (\ref{Composition}) into (\ref{defnrel})
gives
\begin{eqnarray}
  \label{Cond1}
  (Z_\mu)^i&=&{\textstyle\frac\partial{\partial p^\mu_i}}h\\
  \label{Cond2}
  (Z_\mu)^\mu_i&=&-{\textstyle\frac\partial{\partial q^i}}h\\
  \label{Cond3}
  (Z_\mu)_0&=&-{\textstyle\frac\partial{\partial p}}
  +(Z_\mu)^i{\textstyle\frac\partial{\partial q^i}}h
  +(Z_\mu)^\nu_i{\textstyle\frac\partial{\partial p^\nu_i}}h.
\end{eqnarray}
Equation (\ref{Cond2}) does not determine $(Z_\mu)^\nu_i$
completely. This corresponds to the fact that the De Donder-Weyl
equations do determine the change of $\varphi^i$ along the direction
${\scriptstyle\frac\partial{\partial x^\mu}}$ uniquely but do not fix
all components of the change of the polymomenta. Rather, only the combination
$\partial_\mu\pi^\mu_i$ is fixed by the equations.\\
The general solution of (\ref{Cond2}) is hence
\begin{eqnarray}\label{GeneralZmunui}
  (Z_\mu)^\nu_i=-{\textstyle\frac1n}\delta^\nu_\mu
  {\textstyle\frac\partial{\partial q^i}}h+(Z'_\mu)^\nu_i
  \textrm{ with } (Z'_\mu)^\mu_i=0.
\end{eqnarray}
So the vectorfields $Z_\mu$ are given by (\ref{Ansatz}),
(\ref{Cond1}), and (\ref{GeneralZmunui}), as the remaining component
$(Z_\mu)_0$ can be computed from the other ones using (\ref{Cond3}),
\begin{equation}
  (Z_\mu)_0=-{\textstyle\frac\partial{\partial p}}
  +{\textstyle\frac{n-1}n}
  {\textstyle\frac\partial{\partial p^\mu_i}}h
  {\textstyle\frac\partial{\partial q^i}}h
  +{\textstyle\frac\partial{\partial p^\nu_i}}h(Z'_\mu)^\nu_i.
\end{equation}
The following theorem is a corollary of the above observations. A
detailed proof can be found in \cite{PauflerRoemer:2001}.
\begin{theorem}{Theorem}
  \begin{enumerate}
  \item For $h\in\mathcal C\infty(\mathcal P)$ of the form
    \begin{equation}
      h(x^\mu,q^i,p^\mu_i,p)=-\mathcal H(x^\mu,q^i,p^\mu_i)-p
    \end{equation}
    there exist locally associated separable Hamiltonean
    $n$-vectorfields $X_h$ on $\mathcal P$, whose projection onto the
    base $\mathcal M$ has maximal rank $n$.
  \item The integral manifolds of $X_h$ (provided they exist)
    correspond to solutions of the De Donder-Weyl Hamiltonean
    $\mathcal H$.
  \item Let $\varphi$ be a (local) solution of the De Donder-Weyl
    equation (\ref{DWeq}) with De Donder-Weyl Hamiltonean $\mathcal
    H$. Then the tangent spaces along $\varphi$ describe an
    $n$-distribution and hence a separable $n$-vectorfield that is
    Hamiltonean w.r.t. $h$.
  \end{enumerate}
\end{theorem}
Next we shall investigate integrability conditions of the
$n$-vectorfield $X_h$, which will lead us to a Hamilton-Jacobi
formulation of the De Donder-Weyl equations.

First of all, it is important to notice that $X_h$ does not project to
a vectorfield on the extended configuration space $\mathcal E$. This
is due to the fact that the solutions of the equations of motion also
depend on the initial momenta. Pictorially, this is shown in figure 1.
\begin{figure}[h]
  \begin{center}
    \includegraphics[width=4cm]{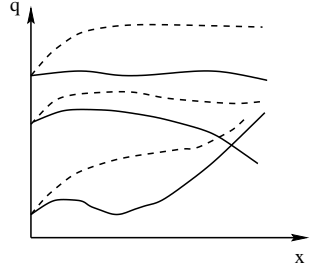}
  \end{center}
  \caption{Projected solutions with different initial momenta.}
\end{figure}

Moreover and unlike the situation in classical mechanics, the totality
of solutions of the De Donder-Weyl equations does not give rise to a
foliation of the multisymplectic phase space. As a matter of fact,
there are many solutions through every point $(x^\mu,q^i,p^\mu_i,p)$
of $\mathcal P$. In general, one can not even expect the existence of
a foliating subset of solutions. A more modest task is the discussion
of foliations of the extended configuration space by solutions of the
De Donder-Weyl equations projected down from $\mathcal P$ to $\mathcal
E$.\\
In classical mechanics, this is achieved by suitable solutions of the
time dependent Hamilton-Jacobi equations
\begin{equation}
  \label{HamJac}
  {\textstyle\frac{\partial S(q,t)}{\partial t}}
    +H(q,{\textstyle\frac{\partial S}{\partial q}},t)=0
\end{equation}
and
\begin{equation}
  \label{HamJac2}
  p_i={\textstyle\frac{\partial S(q,t)}{\partial q^i}},
  \quad
  E=-{\textstyle\frac{\partial S(q,t)}{\partial t}}.
\end{equation}
Equation (\ref{HamJac2}) defines a map
\begin{equation}
  \label{mapT}
  T:\mathcal Q\times  \mathbbm R\rightarrow T^\ast\mathcal Q\times
  \mathbbm R^2,
\end{equation}
where the relation between $T$ and $S$ is given by
\begin{equation}
  \label{T=dS-mech}
  T(t,q^i)=(t,q^i,T_i{\textstyle(t,q^i)},T_0{\textstyle(t,q^i)}),
  \quad
  T_i=\partial_i S,\quad T_0=\partial_t S.
\end{equation}
Analogously, we shall consider a map
\begin{equation}\label{mapT-cov}
  \begin{split}
    T:\mathcal E&\rightarrow \mathcal P\\
    e=(x^\mu,q^i)&\mapsto T(e)=(x^\mu,q^i,T^\mu_i(e)=p^\mu_i,T_0(e)=p)
  \end{split}
\end{equation}
also in field theory.\\
For separable $n$-vectorfields $X_h=Z_1\wedge \cdots Z_n$, we can use
$T$ and the tangential map of the projection onto $\mathcal E$,
$T\pi_{\mathcal E\mathcal P}$, to obtain $n$ vectorfields on $\mathcal
E$,
\begin{equation}
  (\tilde Z_\mu){\textstyle (e)}
  =T\pi_{\mathcal E\mathcal P}Z_\mu\left({\textstyle T(e)}\right).
\end{equation}
The $\tilde Z_\mu$ can be combined to yield a projected distribution
\begin{equation}
  \tilde X_h=\tilde Z_1\wedge \cdots \tilde Z_n,
\end{equation}
which can be expected to be generically integrable and to give a local
foliation of $\mathcal E$.\\
As a matter of fact, a generalisation of the flux straightening
theorem tells us that a (local) foliation of $\mathcal E$ by an $N$ parameter
family of solutions of the De Donder-Weyl equation arises iff these
solutions can be transformed (locally) into constant solutions by a bundle
automorphism $\Phi$ of $\mathcal E$ as indicated in figure 2.
\begin{figure}[h]
  \begin{center}
    \includegraphics[width=8cm]{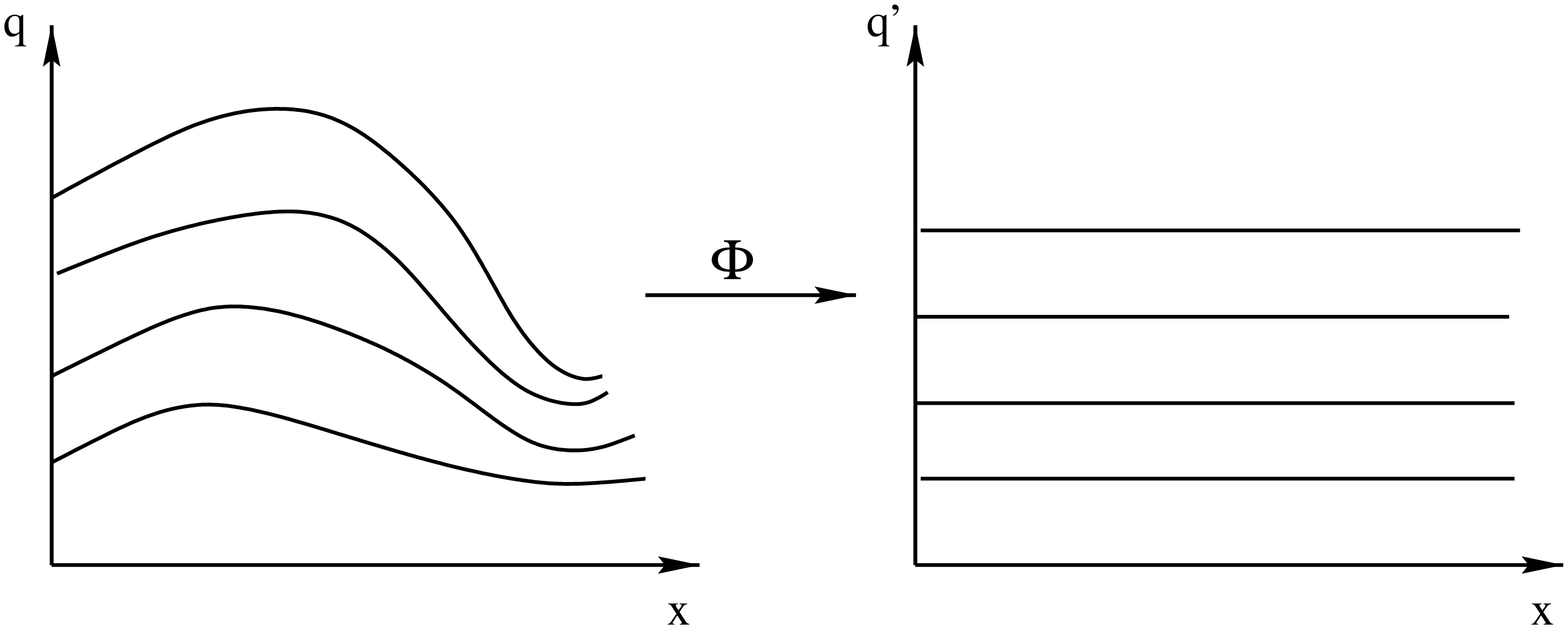}
  \end{center}
  \caption{A foliation yields a trivialisation where the leaves become
    constant sections.}
\end{figure}
More precisely, one can prove the following theorem.
\begin{theorem}
  {Theorem}\label{thm2}
  Let $\mathcal H$ be a regular De Donder-Weyl Hamiltonean on $\mathcal
  P$. Then there exists a local foliation of $\mathcal E$ by projected
  solutions of the De Donder-Weyl equations if and only if there is a (local)
  section $T:\mathcal E\rightarrow \mathcal P$ of the bundle $\mathcal
  P\rightarrow \mathcal E$, which in suitable bundle coordinates
  fulfills
  \begin{eqnarray}
    \label{T1}
    {\textstyle\frac\partial{\partial x^\mu}}T^\mu_i(x,q)
    &=&
    -({\textstyle\frac\partial{\partial q^i}}\mathcal
    H)(x,q,T^\mu_i{\scriptstyle (x,q)})
    \\
    \label{T2}
    {\textstyle\frac\partial{\partial x^\mu}}T_0(x,q)
    &=&
    -({\textstyle\frac\partial{\partial x^\mu}}
    H)(x,q,T^\mu_i{\scriptstyle (x,q)})
    \\
    \label{T3}
    {\textstyle\frac\partial{\partial x^\mu}} T^\mu_i(x,q)
    &=&
    -({\textstyle\frac\partial{\partial q^i}}T_0)(x,q)
    \\
    \label{T4}
    ({\textstyle\frac\partial{\partial q^i}}\mathcal H)
    (x,q,T^\mu_i{\scriptstyle (x,q)})
    &=&
    0
  \end{eqnarray}
\end{theorem}
Here, equations (\ref{T1}) - (\ref{T3}) arise from the De Donder-Weyl
equation on $\mathcal P$, while condition (\ref{T4}) ensures
integrability. A detailed proof can be found in
\cite{PauflerRoemer:2001}.
If a collection of functions $S^\mu(x,q)$, $\mu=1,\ldots,n$,
solves the generalised Hamilton-Jacobi equation
\begin{equation}
  \label{CovHamJac}
  {\textstyle\frac\partial{\partial x^\mu}}S^\mu(x,q)
  +\mathcal H(x,q,{\scriptstyle\frac\partial{\partial q^i}}S^\mu(x,q))
\end{equation}
then
\begin{equation}
  T_i^\mu={\textstyle\frac\partial{\partial q^i}}S^\mu,\quad
  T_0={\textstyle\frac\partial{\partial x^\mu}}S^\mu
\end{equation}
solve the first three conditions (\ref{T1}) - (\ref{T3}) of theorem
\ref{thm2}. Obviously, this constitutes a generalisation of
(\ref{T=dS-mech}).\\
Equation (\ref{T4}) is the foliation condition of the family of
projected solutions. It just means that in suitable bundle coordinates
of $\mathcal E$ the leaves of the foliation constitute a family of
constant fields (sections). The map $T:\mathcal E\rightarrow \mathcal
P$ them arises by a jet prolongation of these fields followed by the
fibre derivative (covariant Legendre transformation)
$\mathbbm F\mathcal L:\mathfrak J_1(\mathcal E)\rightarrow\mathcal P$.

Finally, we will show how (\ref{T1}) - (\ref{T3}) can be cast into an
intrinsic geometric form.\\
Defining the horizontal $(n-1)$-form
\begin{equation}
  \label{DefSform}
  S=S^\mu d^nx_\mu
\end{equation}
and the $2$-horizontal $n$-form $dS$ and using the fact that $\mathcal
P$ can be identified with the space of all $2$-horizontal $n$-forms,
we can write equation (\ref{CovHamJac}) as
\begin{equation}
  \label{CovHamJacGeom}
  h\circ dS=0,
\end{equation}
where $h$ is related to $\mathcal H$ as in (\ref{HamFkt}). The
Cartan-Weyl form
$\Theta$ fulfills,
\begin{equation}
  \label{ThetadS=dS}
  \Theta(dS)=dS.
\end{equation}
More generally, the conditions $T=dS$ and (\ref{CovHamJacGeom}) can be
weakened to
\begin{equation}\label{T1-T3Geom}
  dT=0, \quad
  d(h\circ T)=0,
\end{equation}
where $T$ is a (local)
$2$-horizontal $n$-form $T$ (and hence defines a (local) map
from $\mathcal E$ to $\mathcal P$). The set of equations
(\ref{T1-T3Geom}) is equivalent to the coordinate conditions
(\ref{T1}) - (\ref{T3}).

\end{document}